\documentclass{elsarticle}

\usepackage{makeidx}
\usepackage{wrapfig, color}
\usepackage{verbatim}
\usepackage{algorithm}
\usepackage{algpseudocode}
\usepackage{subfigure}
\usepackage{float}

\usepackage{graphicx}
\usepackage{epstopdf}
\usepackage{multirow}
\usepackage{amsmath}
\usepackage[nomargin,inline,marginclue,draft]{fixme}

\usepackage{mathtools}

\usepackage{listings}

\def\x{{\mathbf x}}

\begin{document}
\begin{frontmatter}

\title{Data-Aware Approximate Workflow Scheduling}

\author{Dengpan~Yin}
  \ead{yindengpan@gmail.com}
\address{F5 Networks Inc.\\ 401 Elliott Avenue West\\ Seattle, WA 98119-4017, USA}

\author{Tevfik~Kosar}
  \ead{tkosar@buffalo.edu}
\address{Department of Computer Science and Engineering\\ University at Buffalo (SUNY)\\ Buffalo, New York, 14260, USA}

\pagebreak
\begin{abstract}
Optimization of data placement in complex scientific workflows has become very crucial since the large amounts of data generated by these workflows significantly increases the turnaround time of the end-to-end application. It is almost impossible to make an optimal scheduling for the end-to-end workflow without considering the intermediate data movement. In order to reduce the complexity of the workflow-scheduling problem, most of the existing work constrains the problem space by some unrealistic assumptions, which result in non-optimal scheduling in practice. In this study, we propose a genetic data-aware algorithm for the end-to-end workflow scheduling problem. Distinct from the past research, we develop a novel data-aware evaluation function for each chromosome, a common augmenting crossover operator and a simple but effective mutation operator. Our experiments on different workflow structures show that the proposed GA based approach gives a scheduling close to the optimal one.

\end{abstract}


\end{frontmatter}
\newpage

\section{Introduction}
The increasing size and complexity of scientific applications require more advanced workflow management and scheduling techniques, which take the data dependencies and the cost of data movement into account. A workflow may comprise many small tasks, of which the independent tasks can be dispatched to different computation sites in the distributed environment and executed in parallel. End-to-end workflows which require a lot of data handling, i.e. `data-intensive workflows, have been applied in many fields such as astronomy\cite{dataintensive_astronomy}, bioinformatics\cite{dataintensive_bio} and high-energy physics\cite{dataintensive_phy}.  In these applications,  terabytes of data will be processed by the workflow and some important intermediate data need to be stored for future use.  Workflow scheduling problem has been studied for decades; however, to our best knowledge, none of the existing algorithms can give an optimal solution for the data intensive workflows.  Along with the increasing data size in the workflow applications, it is imperative for us to develop a new algorithm to address this scheduling problem with a close to optimal approach. 

The workflow scheduling problem has been well studied for many years and today it is still a very active research area. In order to minimize the turnaround time, many approximation algorithms have been proposed, such as genetic algorithms \cite{lee:generic,wu:ga,zom:ga}, simulated annealing algorithms \cite{yong:sa} and ant colony algorithms \cite{wei:ant}. Although these algorithms do not guarantee an optimal solution, they guarantee to generate an acceptable solution in a timely manner. The quality of the solution is controlled by a series of parameters.   

There are also several approaches that aim to find the optimal solution of the workflow scheduling problem. Chou and Chung proposed to find the optimal scheduling for workflows on multiprocessors~\cite{hong:opt}. However, the communication cost is ignored due to low latency between processors. In \cite{chang:opt}, Chang and Jiang presented a state space search algorithm to address the problem. They used the critical path length as an underestimate of the actual cost function to guide the expansion of the state during the solution exploring process. In their research, the communication cost between tasks is also ignored. Kwok and Ahmad~\cite{kwok:state} proposed a parallel state space search approach based on A-Star algorithm. They applied state-pruning techniques to reduce the search space, and assumed that the bandwidth between different processors are homogeneous.

Wang et al~\cite{wang} proposed a state space search approach based on A-star algorithm to solve the workflow scheduling problem in distributed environments. In their research, they assumed that the computation power of the computational sites and the network bandwidth between them are heterogeneous. They claimed that their solution to be optimal. Lin \cite{Lin} studied the same problem and pointed out that Wang's solution is not optimal in some cases. Wang et al assumed that one of the immediately preceding communications of a task must be performed just before the execution of the task. Lin relaxed this constraint and rearranged the task execution and task communication orders in order to get the optimal scheduling. 

Both Lin and Wang et al have assumed that the computation sites should be in either execution state or communication state. It cannot overlap execution and data transfer. With this constraint, the problem can be simplified, however, the scheduling turns out to be non-optimal in some cases. It makes sense that a computation site can receive input data for a task when it is executing another task. The data will be ready at the execution finish time of the previous task. Based on this investigation, we further remove that constraint and decrease the turnaround time by overlapping the execution and data placement. 

There is plethora of work on optimizing the speed of data transfer between any two nodes in an end-to-end workflow~\cite{TCC_2016,Cluster_2015,IGI_2012,Grid_2008,DADC_2008, WORLDS_2004,ScienceCloud_2013,Thesis_2005,NDM_2011,Royal_2011,NDM_2012,JGrid_2012,DADC_2009}. But these work do not consider co-scheduling of compute and data placement tasks, and could be used in conjunction with a separate data-aware workflow scheduler to reduce the overall data transfer time during workflow execution. 

In all of these previous works mentioned above, little attention is paid to the optimization of data placement in end-to-end workflow scheduling. None of these work mention about how to deal with the scheduling problem when there are extra data that need to be staged-in from remote site or to be staged-out to a remote site for a task. 
In this paper, we present a genetic algorithm approach for the end-to-end workflow scheduling problem. Distinct from the past research, we develop a novel data-aware evaluation function for each chromosome, a common augmenting crossover operator and a simple but effective mutation operator. We also present a stage-in data location optimization algorithm.


\begin{figure}[t]
\begin{center}
\includegraphics[scale=0.6]{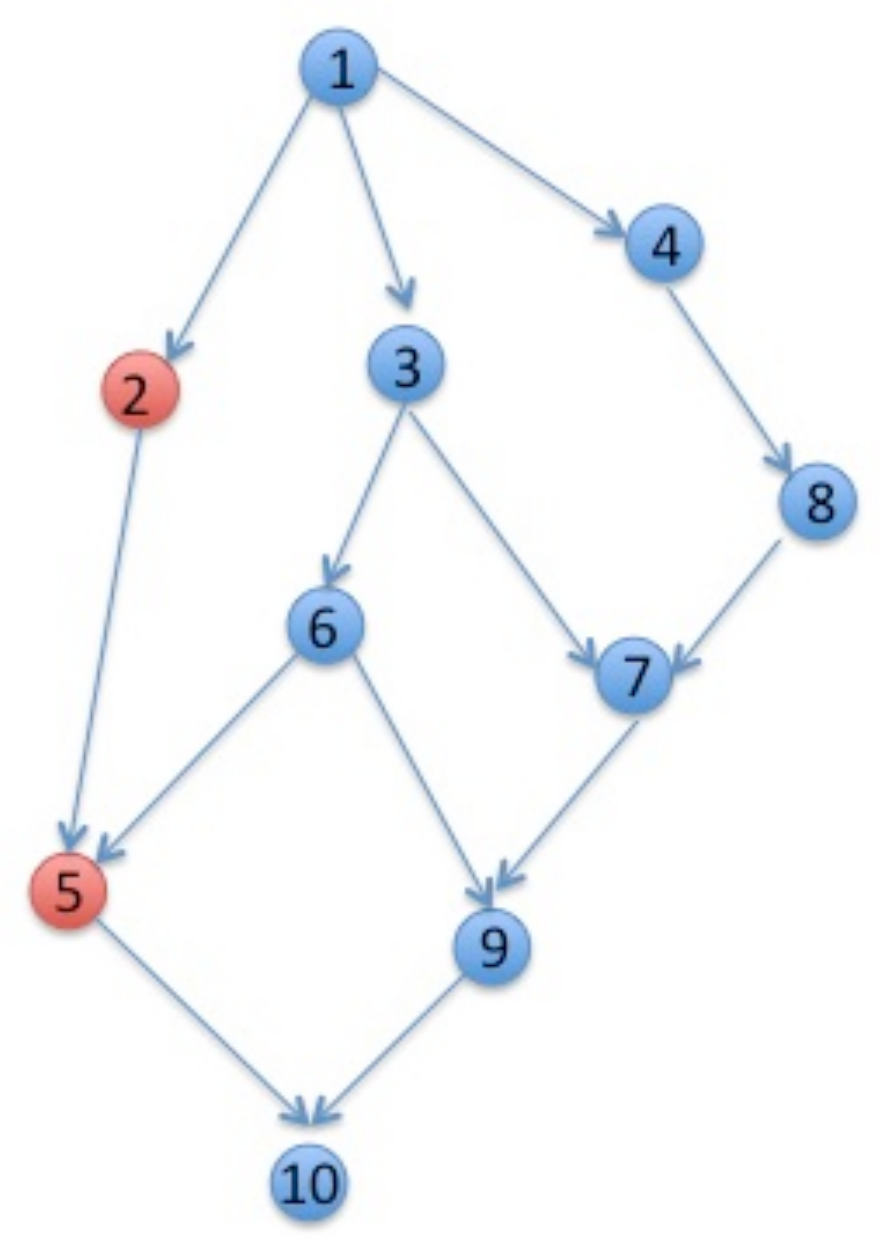}
\caption{Sample workflow graph representation}
\label{fig:taskheight}
\end{center}
\end{figure}

\section{Task Height \& Chromosome Enconding}

To facilitate the crossover and mutation operation, we define the task height similar to the definition in \cite{GA_Zomaya, GA_Tsujimura, GA_Hou}. {

\begin{equation}
Height(t_i) =  \left \{
\begin{array}{lr}
0 & if~t_i~is~root \\
\underset{ t_j \in pred(t_i)} {\operatorname{max}} Height(t_j) + 1 & otherwise 
\end{array}
\right .
\end{equation}

From the definition, It is easy to see that if $t_j$ is an ancestor of $t_i$, then $Height(t_j) < Height(t_i)$. It is always feasible that a task with a smaller height is executed before a task with a larger height and tasks with the same height are executed in an arbitrary order.  However, there are some occasions that it is also feasible to execute a task with a larger height before a task with a smaller height. For example,  in Figure \ref{fig:taskheight}, the height of each task is shown in Table \ref{tb:taskheight}. The height of $t_2$ is 1, the height of $t_6$, $t_7$, $t_8$ and $t_9$ are greater than one, yet it is still feasible to execute $t_1$ after them. 

In order to have a uniform relationship between task execution order and task height, we introduce two additional height notations, $height_{eq}$ and $height_{soft}$. The height of a task is calculated topdown and the equivalent height is calculated bottom up. A soft height is defined as a value in between the original height and equivalent height.  $Height_{soft}$ is the same as $Height'$ defined in \cite{GA_Tsujimura}. The height and equivalent height of a task are fixed, while the soft height can be different for each chromosome. Since each chromosome will have a distinct task height array and the task execution orders are based on the soft height, when we have a large number of chromosomes, we can explore a broad search space for task execution orders and increase the probability of finding an optimal scheduling.

\begin{equation}
\label{eq:heighteq}
Height_{eq}(t_i) =  \left \{
\begin{array}{lr}
Height(t_i) & if~t_i~is~end~task \\
\underset{ t_j \in Suc(t_i)} {\operatorname{min}} Height_{eq}(t_j)  - 1 & otherwise 
\end{array}
\right .
\end{equation}

\begin{equation}
\label{eq:heightsoft}
Height_{soft}(t_i) = Height(t_i) + rand() \% (Height_{eq}(t_i) - Height(t_i) + 1) 
\end{equation}

A chromosome should contain the information of solution which it represents. For the workflow scheduling problem, the solution is a complete scheduling, a mapping from each task to a processor.  For a complete scheduling, each task in the workflow will be assigned to one processor in a certain order. The tasks on each process will be execute sequentially and the tasks on different processors will be executed in parallel. To represent the information of a particular scheduling, the processor on which the tasks will be executed and the execution order on each processor should be considered. 

In this study, we have the following assumptions for processor IDs and task IDs:
\begin{enumerate}
\item The start task ID is 1 and it is a dummy task with zero execution time and zero data transfer to its successors. This will ensure the entire workflow will have a single start point.
\item The end task is a dummy task with zero execution time and zero data transfer between its predecessors.  This will ensure the entire workflow will have a single end point.
\item The processor ID is an integer and greater than zero.  The will enable us to user zero as a delimiter during encoding.  
\end{enumerate}

Since both processor ID and task ID are integer, it is very convenient for us to use integer representation during encoding. We append the ID of each task assigned to a processor to the processor ID and use zero as a delimiter for the task assignment for each processor. For example, the encoding for the scheduling in Table \ref{tb:sched} is: {\small 0 1 2 5 0 2 3 6 9 0 3 4 8 7.}

\begin{table}[t]
\centering
\caption{Task height for the sample workflow}
\centering
\begin{tabular} {|c| c| c| c| c| c| c| c| c| c| c|}
\hline
~& $t_1$ & $t_2$ & $t_3$ & $t_4$ & $t_5$ & $t_6$ & $t_7$ & $t_8$ & $t_9$ & $t_{10}$ \\
\hline
Height & 0 & 1 & 1 & 1 & 3 & 2 & 3 & 2 & 4 & 5 \\
\hline
$Height_{eq}$ & 0 & 3 & 2 & 1 & 4 & 3 & 3 & 2 & 4 & 5 \\
\hline
$Height_{soft}$ & 0 & 2 & 1 & 1 & 3 & 3 & 3 & 2 & 4 & 5 \\
\hline
\end{tabular}
\label{tb:taskheight}
\end{table}

\begin{table}[h]
\centering
\caption{A complete scheduling}
\centering
\begin{tabular}{|c|l|}
\hline
processor & task \\
\hline
$P_1$ & 2 ~ 5 \\
\hline
$P_2$ & 3 ~ 6 ~ 9 \\
\hline
$P_3$ & 4 ~ 8 ~ 7 \\
\hline
\end{tabular}
\label{tb:sched}
\end{table}  
}

\section{Population \& Crossover Operator}
The population is a set of chromosomes. To generate a chromosome, first calculate the soft height for each task, and then assign tasks in order of the soft height randomly to a processor. Each chromosome will maintain its own task soft height array and will govern the mutation process for each chromosome.  The process of generating the population is shown in Algorithm \ref{alg:genpop}.


\begin{algorithm} [h]
\caption{Generate population}
\label{alg:genpop}
\begin{algorithmic}[1]
\For {$i = 1$ to $population ~ size$}
	\State calculate the soft height of each task
	\State maxHeight $\leftarrow$ the maximal soft height of the tasks for the current chromosome
	\For {$j = 1$ to $maxHeight$ }
		\ForAll {$task$ such that $Height_{soft}(task) = j$} 
			\State randomly assign $task$ to a $processor$
		\EndFor
	\EndFor
	\State encoding the complete scheduling to a $chromosome$
	\State add the $chromosome$ to the $population$
\EndFor
\end{algorithmic}
\end{algorithm}

The crossover operation is performed on two chromosomes with the hope of generating new offsprings with higher fitness. The genes in the conventional chromosome are independent to each other, which makes the crossover operation simple and feasible. Taking two feasible chromosomes and intermingling their genes will still generate two feasible chromosomes. However, in the workflow scheduling problem, it is more complex. The chromosome is comprised of a set of genes representing task assignment, processor ID and delimiter. Each chromosome represents a complete scheduling and should cover all the task assignments.   Simply exchanging gene segments is prone to result in incomplete chromosome and gene duplication. For example, suppose we have the following two chromosomes:\\
{\small
\indent chrom1 = 0 1 2 5 0 2 3 6 9 0 3 4 8 7 \\
\indent chrom2 = 0 1 2 9 0 2 3 6 7 0 3 4 8 5 \\ 
}
After exchanging the first 4 integers of these two chromosomes, we get the following two offsprings:\\
{\small
\indent offspring1 = 0 1 2 9 0 2 3 6 9 0 3 4 8 7 \\
\indent offspring2 = 0 1 2 5 0 2 3 6 7 0 3 4 8 5 \\ 
}
It is easy to observe that $offspring1$ is incomplete since it lacks the information of where 5 is assigned. Meanwhile, it has duplicate and inconsistent information since it indicates that 9 is assigned to both processor 1 and 2. $offspring2$ has the same problem as $offspring1$.

Past research has been done to perform crossover to tasks assigned to each processor, do some adjustment on the fly and finally output two feasible offsprings.  This approach is little complex for implementation. Besides, there is no clear evidence showing that such a complex crossover  outperforms a simple operation by replacing some chromosomes of small fitness values with some new fresh chromosomes.

In this work, a new approach is introduced to simplify the crossover operation. In our approach, instead of replacing two parents with two new offsprings,  the population will take one new offspring generated from two parents as well as keeping the parents. In the end, the population size will increase to one and a half times of the original size. A ranking selection method can reduce the population back to its original size.
If a task is assigned to the same processor in the parent chromosome, it will remain this assignment in the offspring. The tasks having different assignment in the parent chromosome will have a random assignment in the offspring. For example, an offspring could be {\small $offspring$ = 0 1 2 9 0 2 3 6 5 0 3 4 8 7} if the parents are $chromosome1$ and $chromosome2$.  The details of the operation are shown in Algorithms \ref{alg:crossover} and \ref{alg:genoffspring}. 

\begin{algorithm}[h]
\caption{Crossover}
\label{alg:crossover}
\begin{algorithmic}[1]
\ForAll {$(chom_i, chrom_j)$ such that $(chrom_i \in population) \wedge (chrom_j \in population) \wedge (chrom_i \notin p) \wedge (chrom_j \notin p$)}
\State $offspring \leftarrow GenerateOffspring(chrom_i, chrom_j)$
\State add $offspring, chrom_i, chrom_j$ to $p$
\EndFor
\State $population$ $\leftarrow$ select the best $population size$ chromosomes from $p$.
\end{algorithmic}
\end{algorithm}

\begin{algorithm}[t]
\caption{GenerateOffspring($chrom_i, chrom_j$)}
\label{alg:genoffspring}
\begin{algorithmic}[1]
\State $taskToProcessor[taskCount] \leftarrow 0$
\ForAll {$t$ such that $t$ is assigned to the same processor in both $chrom_i$ and $chrom_j$}
	\State $taskToProcessor[t] \leftarrow processorID$
\EndFor
\State randomly select the soft height of each task from parents
\State maxHeight $\leftarrow$ the maximal soft height of the tasks
\For {$j = 1$ to $maxHeight$ }
	\ForAll {$t$ such that $Height_{soft}(t) = j$} 
		\If {$taskToProcessor[t] \ne 0$}
			\State assign $t$ to $taskToProcessor[t]$
		\Else
			\State randomly assign $t$ to a $processor$
		\EndIf
	\EndFor
\EndFor
\State encoding the complete scheduling to a $chromosome$
\end{algorithmic}
\end{algorithm}

\begin{algorithm}[t!]
\caption{Mutation}
\label{alg:mutation}
\begin{algorithmic}[1]
\State randomly choose two processors $i$ and $j$.
\State perform a merge sort to the tasks assigned to processor $i$ and $j$ according to their soft height
\ForAll {task in the sorted task array}
\State randomly assign this task to processor $i$ or $j$
\EndFor
\end{algorithmic}
\end{algorithm}

\section{Mutation Operator \& Fitness Function}

The mutation operation in the workflow scheduling problem should ensure the chromosome after mutation to be a complete scheduling without redundancy. Besides, the task execution order should follow the data dependency constraints. Simply changing a task ID will cause the chromosome to contain duplicate and inconsistent information. Moreover, the chromosome after this mutation is incomplete since it lacks the task assignment for the task before mutation.  Randomly exchange the position of two tasks might violate the data dependency constraints. 

In this work, a simple and effective mutation method is introduced.  First, we randomly choose two processors and perform a merge sort to the tasks according to their soft height, and then we randomly assign each task to these two processors sequentially. Algorithm \ref{alg:mutation} shows the detail of the mutation operation.
The fitness is calculated based on the turnaround time of the scheduling represented by the chromosome. Since the objective is to find a scheduling with the shortest turnaround time, a chromosome with shorter turnaround time will have higher fitness value. The fitness in this work is calculated by the difference of the maximal turnaround time in the population and the turnaround time of the current chromosome. 

To get the turnaround time of scheduling represented by the chromosome, the task assignment and execution order must be retrieved from the chromosome. The tasks assignment on each processor is already sorted according to the soft height. Performing a merge sort to the task assignments on each processor will output an array of task assignments in nondecreasing order of soft height, as shown in Algorithm \ref{alg:sched}.  Preliminary results for this data-aware genetic algorithm were presented in ~\cite{ICECCO13}

\begin{algorithm}[h]
\caption{RetrieveSched}
\label{alg:sched}
\begin{algorithmic}[1]
\State $sched \leftarrow$ task assignment on processor 1
\For { $pid = 2$ to $processor ~ count$}
	\State merge tasks assigned to processor $pid$ with $sched$ according to softheight
\EndFor
\ForAll {task in $sched$}
\State calculate the earliest stage-in time 
\State calculate the earliest start time
\State calculate the earliest time to transfer data
\State calculate the earliest stage-out time
\EndFor
\ForAll {processor}
\State {calculate the finish time of the last task assigned to this processor}
\State {calculate the stage-out finish time time of each task which has data to be staged-out}
\EndFor
\State return the maximum of latest execution finish time and latest stage-out finish time
\end{algorithmic}
\end{algorithm}  

\section{Stage-in Data Placement Optimization}

When there is a large data set in the data intensive workflow, there could be many mappings from data set to storage site. Figure \ref{fig:stagein_opt} shows one possible mapping from tasks to computing site and from stage in data to storage site. A poor task scheduling will increase the turnaround time of the workflow, resulting in an inefficient utilization of the computing resource. Similarly, a poor mapping from stage in site to storage site will degrade the overall performance since it will require a longer time to deliver the data to the computing site. 

To formalize the stage in data location optimization problem, we assume there are a number of storage sties and computing sites  connected by homogeneous or heterogeneous network
There are many data files with different size to be delivered between the storage site and the computing site. Each file a fixed destination. The question is where to put each data file such that the deliver time for all the data files is minimized.    

\begin{figure}[h]
\begin{center}
\includegraphics[scale=0.5]{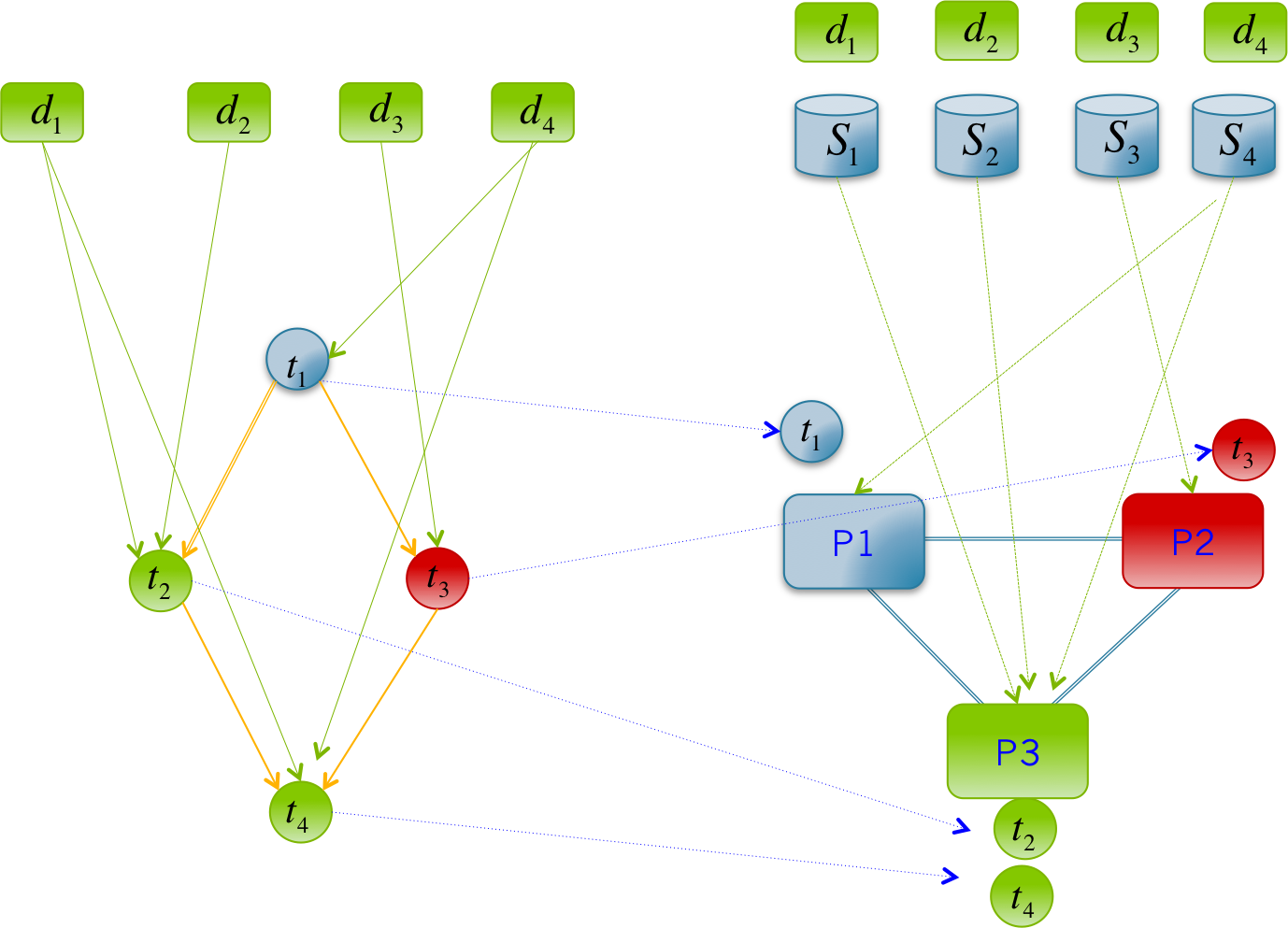}
\caption{Mapping data files to storage sites}
\label{fig:stagein_opt}
\end{center}
\end{figure}

It is impossible to find an true optimal solution in polynomial time. In this research, a close to optimal algorithm is proposed. The main idea is to distribute data to each storage site proportional to the bandwidth between the storage site and the computing site. Make the data amount distributed to each storage site as close as possible to the desired portion. Make a replica of each data file when  conflict  occurs for placing a particular data file which is to be delivered to multiple computing site. In this case,  this data file is the input for multiple tasks.  Algorithm \ref{alg:sin_opt} describes the details to address the stage in data location optimization problem. 

\begin{algorithm}[h]
\caption{Stage in data location optimization}
\label{alg:sin_opt}
\begin{algorithmic}[1]
\ForAll {Computing site $P_i$ that has data to be staged in}
\State  Sort all data files to be delivered to $P_i$ in descending order
\State  Sort the bandwidth from all storage sites to  $P_i$ in descending order
\ForAll {Storage site with connection to the computing site}
\State  Distribute data to the storage site with a higher bandwidth
\State  Find the largest $k$ such that the sum of the first $k$ data are less than or equal to the desired data amount 
\State  Distribute them to the corresponding storage site
\State From the rest of the data, binary search the data closest to the difference between the portioning data and the sum of the first $k$ data 
\State Distribute the data to the corresponding site
\EndFor
\EndFor
\end{algorithmic}
\end{algorithm}

\section{Evaluation}

The transfer time lower bound for the data to be delivered to a computing site is the quotient of the summation of the data amount and the summation of the bandwidth from each storage site to the computing site.  The transfer time lower bound for all these computing sites is the maximum among them.  Figure \ref{fig:stagein_opt} shows the experimental results for different number of data files. The transfer time is surprisingly  close to the lower bound for both homogeneous and heterogeneous distributed systems. 

\begin{figure}[h]
\begin{center}
\includegraphics[scale=0.31]{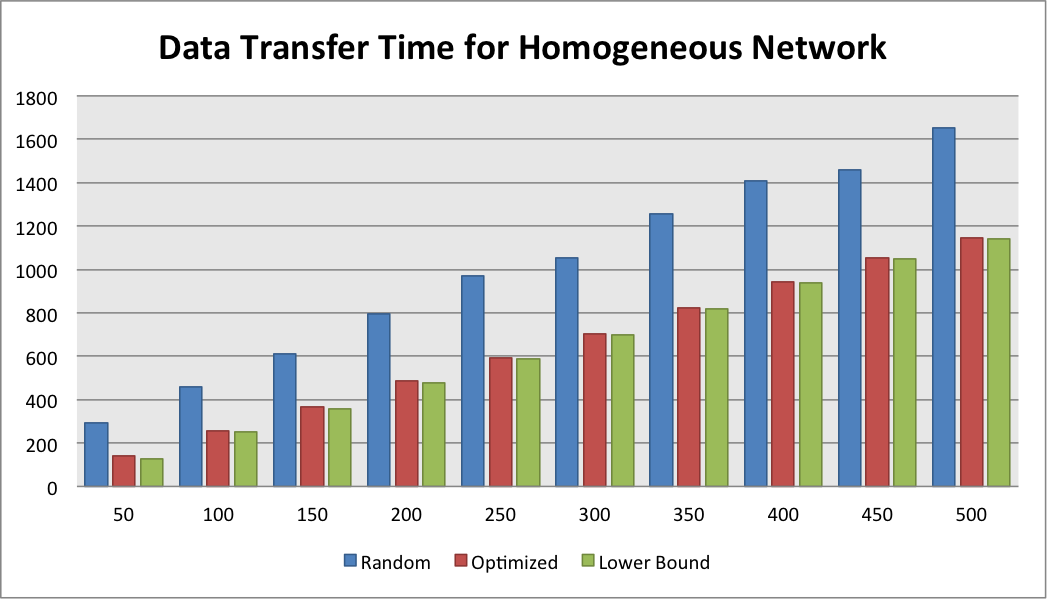}
\includegraphics[scale=0.30]{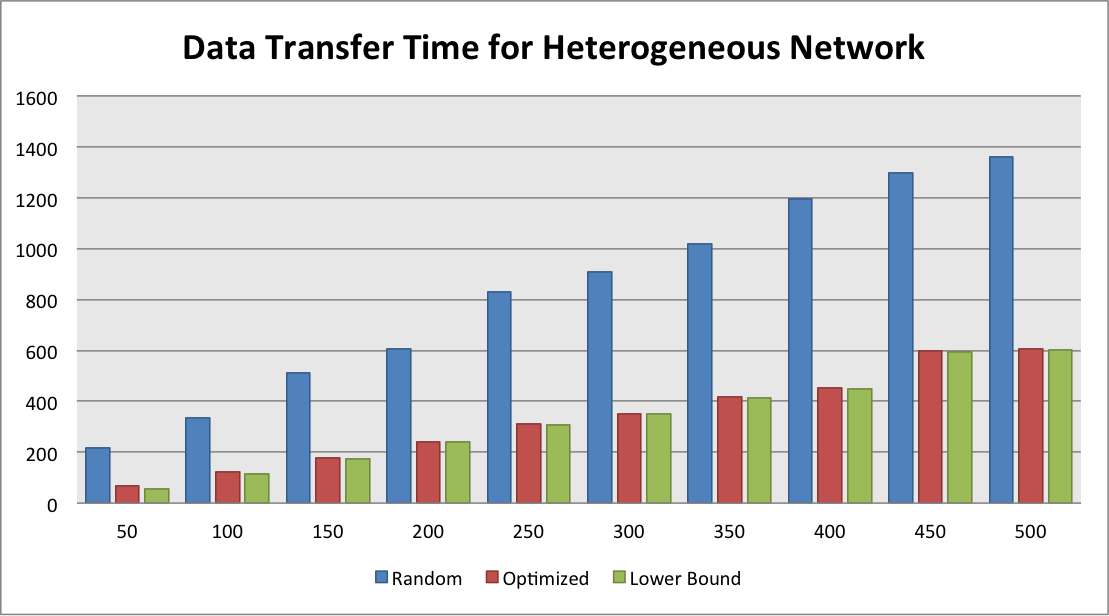}
\caption{Optimizing the stage-in data location for homogeneous and heterogeneous distributed systems}
\label{fig:stagein_opt}
\end{center}
\end{figure}

\newpage

\begin{figure}[h]
\begin{center}
\includegraphics[width=0.80\textwidth]{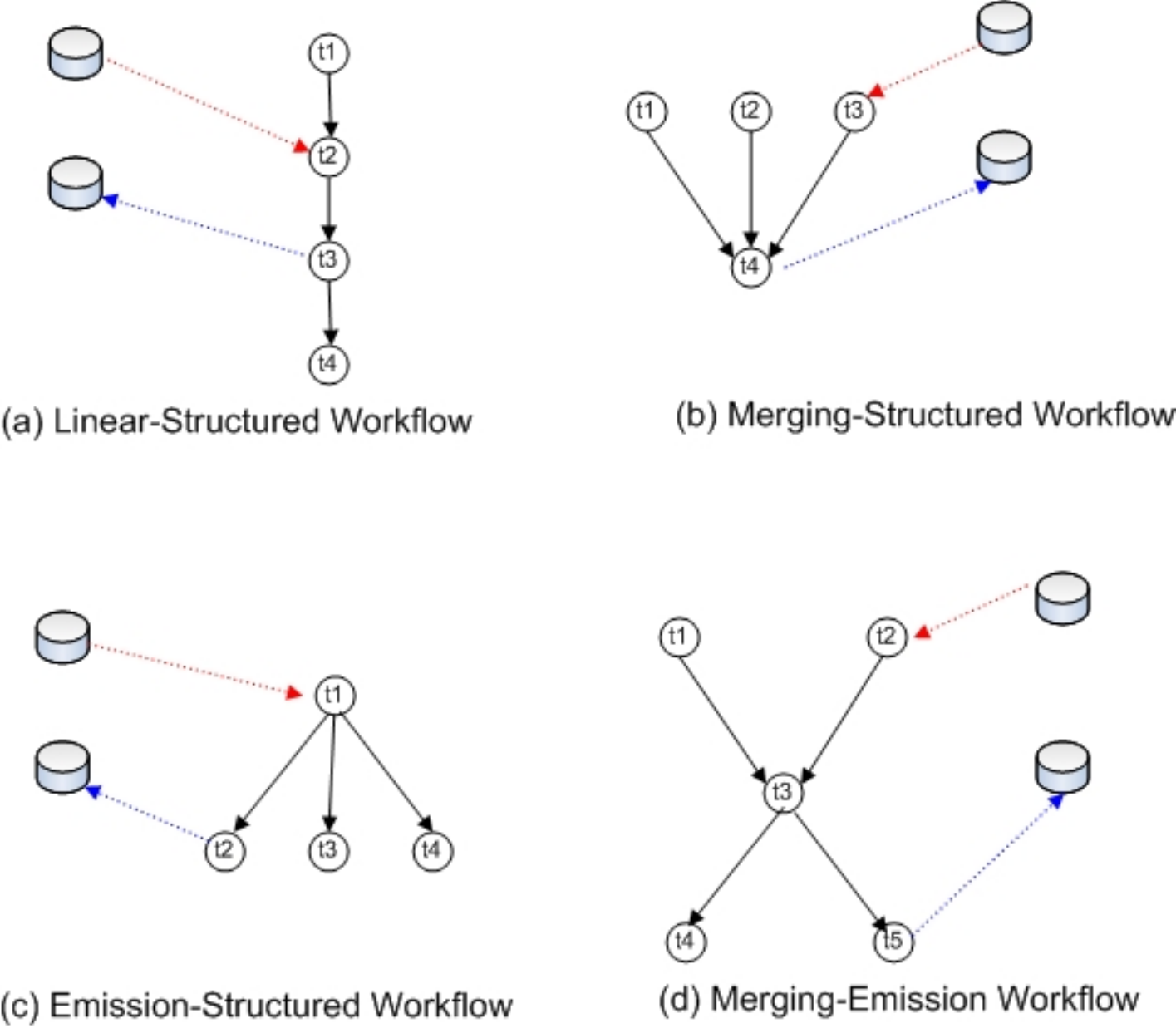}
\caption{Typical workflow structures}
\label{fig:workflow_structure}
\end{center}
\end{figure}

We have tested our GA based workflow optimization algorithm on four different common workflow structures: {\em i)} linear-structured workflow; {\em ii)} merging-structure workflow; {\em iii)} emission-structured workflow; and {\em iv)} merging-emission workflow, as illustrated in Figure \ref{fig:workflow_structure}. 

\begin{figure}[t]
 \begin{center}
 \begin{tabular}{cc}
 \resizebox{66mm}{!}{\includegraphics{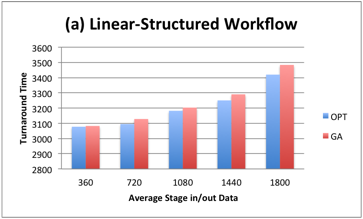}} 
 \resizebox{66mm}{!}{\includegraphics{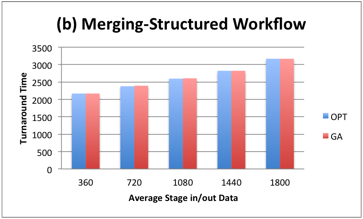}} \\
 \resizebox{66mm}{!}{\includegraphics{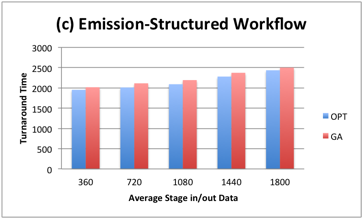}} 
 \resizebox{66mm}{!}{\includegraphics{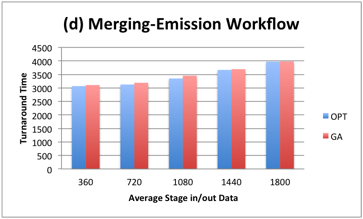}} \\
 \end{tabular}
 \caption{Turnaround time comparison of  GA and optimal scheduling}
 \label{fig:time_typical_workflow}
 \end{center}
\end{figure}

Then we compared the performance of our GA based algorithm to the theoretical optimal case (OPT). In all four workflow structures, our GA based algorithm performed very close to the optimal case. The results are shown in Figure~\ref{fig:time_typical_workflow}.

\section{Conclusion}

We have presented a genetic data-aware algorithm for the end-to-end workflow scheduling problem. Our proposed GA based algorithm differs from the previous research in two significant ways. First, it is data-aware. The fitness function considers all the data involved in the workflow, the stage-in data, stage out data, and the intermediate data. Second, the mutation and crossover operation are simple but effective. Our experiment results show that the proposed GA based approach gives a scheduling close to the optimal one.

\bibliographystyle{elsarticle-num}
\bibliography{didc,thesis}

\end{thebibliography}
\end{document}